\documentclass[conference]{IEEEtran}
\IEEEoverridecommandlockouts
\usepackage{cite}
\usepackage{amsmath,amssymb,amsfonts}
\usepackage{algorithmic}
\usepackage{graphicx}
\usepackage{textcomp}
\usepackage{xcolor}
\def\BibTeX{{\rm B\kern-.05em{\sc i\kern-.025em b}\kern-.08em
    T\kern-.1667em\lower.7ex\hbox{E}\kern-.125emX}}
\usepackage{tikz}
\usetikzlibrary{shapes.geometric, arrows}
\usepackage{graphicx}
\usepackage{float}
\usepackage[colorlinks=true,linkcolor=blue,urlcolor=blue,citecolor=blue]
{hyperref}

\title{A Recall-First CNN for Sleep Apnea Screening from Snoring Audio}

\makeatletter
\newcommand{\linebreakand}{%
  \end{@IEEEauthorhalign}
  \hfill\mbox{}\par
  \mbox{}\hfill\begin{@IEEEauthorhalign}
}
\makeatother

\author{\IEEEauthorblockN{Anushka Mallick}
\IEEEauthorblockA{\textit{Sai International School}\\
Bhubaneswar,India\\
anushkamallick475@gmail.com
}
\and
\IEEEauthorblockN{Afiya Noorain}
\IEEEauthorblockA{\textit{Sai International School}\\
Bhubaneswar,India\\
afiyanoorain296@gmail.com}
\and
\IEEEauthorblockN{Ashwin Menon}
\IEEEauthorblockA{\textit{Gems New Millennium School}\\
Dubai,United Arab Emirates\\
 ashwinmenon2108@gmail.com}
\linebreakand
\IEEEauthorblockN{Ashita Solanki}
\IEEEauthorblockA{\textit{Queen's Valley School}\\
Dwarka,India\\
ashitasolanki11@gmail.com}
\\
\and
\IEEEauthorblockN{Keertan Balaji}
\IEEEauthorblockA{\textit{ Department of Computer Science and Engineering}\\
\textit{ S.R.M Institute of Science and Technology}\\
 Chennai, India\\
 kb9379@srmist.edu.in
}}

\begin{document}
\maketitle
\begin{abstract}
Sleep apnea is a serious sleep-related breathing disorder that is common and can impact health if left untreated. Currently the traditional method for screening and diagnosis is overnight polysomnography. Polysomnography is expensive and takes a lot of time, and is not practical for screening large groups of people. In this paper, we explored a more accessible option, using respiratory audio recordings to spot signs of apnea.We utilized 18 audio files.The approach involved converting breathing sounds into spectrograms, balancing the dataset by oversampling apnea segments, and applying class weights to reduce bias toward the majority class. The model reached a recall of 90.55 for apnea detection. Intentionally, prioritizing catching apnea events over general accuracy. Despite low precision,the high recall suggests potential as a low-cost screening tool that could be used at home or in basic clinical setups, potentially helping identify at-risk individuals much earlier.
\end{abstract}

\section{Introduction}
Sleep apnea is a disorder where breathing repeatedly slows or stops during sleep. These pauses reduce oxygen saturation and disrupt normal sleep cycles, which over time contribute to hypertension, metabolic dysfunction, cognitive impairment, and daytime fatigue. Prevalence estimates vary across studies, but a consistent concern is that many cases remain undiagnosed for years, diminishing the effectiveness of treatment once it begins.  

Polysomnography (PSG)\cite{POLY} remains the diagnostic standard, as it records airflow, oxygen levels, brain activity, and other physiological signals. However, PSG is costly, requires trained staff, and typically involves an overnight stay in a sleep laboratory. These requirements limit its practicality for large-scale screening or routine home monitoring.  

Researchers have therefore explored alternatives such as oxygen monitoring, heart activity, or mandibular movement. Respiratory audio is a particularly promising option because it is simple to record and widely accessible. Yet audio-based approaches face challenges: real-world recordings are noisy, apnea events are relatively rare, and most existing models show low recall, often missing true apnea cases. This is a critical weakness in a screening context where sensitivity is more important than overall accuracy.  

In this study, we test whether deep learning applied directly to respiratory audio can achieve recall-first performance. Using recordings from the MIT-BIH Polysomnographic Database\cite{dataset}, we convert audio into spectrograms and train a convolutional neural network (CNN) with recall as the primary objective. The design emphasizes minimizing false negatives, accepting lower precision if it ensures nearly all apnea cases are detected.  

The contributions of this study are as follows:
\begin{itemize}
    \item We train a ResNet-based\cite{resnet} CNN directly on spectrograms of breathing sounds, avoiding reliance on hand-crafted features\cite{HANDCRAFTED} or additional sensors.
    \item We address the imbalance between apnea and non-apnea events through oversampling and class weighting, preventing the model from ignoring rare events.
    \item We demonstrate that a recall-first approach can achieve sensitivity above 90\%, suggesting that audio alone may serve as a low-cost and device-independent screening tool.
\end{itemize}

Compared with earlier work, our approach shifts the focus. Traditional audio-based models struggled with noise and low sensitivity, while feature-driven machine learning relied heavily on patient data and often failed to generalize. Wearable deep learning systems achieved strong performance but required specialized devices. In contrast, our results show that audio alone, when paired with a recall-first strategy, can provide a scalable and accessible option for early apnea screening.

\begin{table*}[!htbp]
\centering
\caption{COMPARATIVE SUMMARY OF APPROACHES}
\label{tab1}
\resizebox{\textwidth}{!}{%
        \begin{tabular}{|l|c|c|c|l|}
            \hline
            \textbf{Method} & \textbf{Dataset Type} & \textbf{Sensitivity} & \textbf{Specificity} & \textbf{Key Gaps} \\
            \hline
            Classical (LDA)\cite{automatic_classification_of_subjects} & Acoustic snoring signals & ~90\% & ~41\% & High false positives, noise sensitivity, lacks robustness \\
            \hline
            Machine Learning\cite{jmir_2022} & Demographic + snoring features & ~85–90\% & ~50–60\% & Feature reliance, limited generalization, small datasets \\
            \hline
            Deep Learning (Wearable)\cite{real_time}\cite{Decoding_Sleep} & Vibration/snoring signals & ~92\% & ~70\% & Computational cost, interpretability challenges, dataset size \\
            \hline
            Domain-Specific (MM)\cite{pepin}\cite{JANOTT2018106} & Mandibular movement signals & ~95\% & ~85\% & Limited external validation, needs broader trials \\
            \hline
            Scoping Reviews\cite{IOT_Healthcare}\cite{Design_of_embedded_real-time_system} & Multiple datasets & Varies & Varies & Dataset imbalance, inconsistent metrics, reporting gaps \\
            \hline
        \end{tabular}
    }
    \label{tab:comparative_summary}
\end{table*}

\section{LITERATURE REVIEW}

Over the years, a wide range of methods have been explored for detecting sleep apnea, each contributing important progress in its own way. The earliest attempts relied on handcrafted acoustic features combined with classical statistical tools such as Linear Discriminant Analysis (LDA). These systems achieved promising sensitivity levels of around 90\%, showing they could correctly identify most apnea cases. However, their specificity was much lower, at only 41\%, which meant that many healthy individuals were misclassified as having apnea \cite{automatic_classification_of_subjects}. Building on these foundations, machine learning techniques like Random Forests and other non-linear classifiers were introduced. By incorporating acoustic features alongside simple patient data such as BMI and neck circumference, they offered a more flexible approach to screening moderate-to-severe cases across larger populations. Their sensitivity remained consistently high (about 85–90\%) and specificity improved somewhat, but these models still leaned heavily on carefully designed features and often struggled to adapt to different datasets important\cite{jmir_2022}.

The rise of deep learning brought a major shift, allowing models to learn directly from raw or minimally processed data rather than depending on feature engineering. For example, one study using a neck-wearable piezoelectric sensor showed that deep learning could distinguish severe sleep apnea from habitual snoring with greater accuracy than earlier acoustic-only methods. Similarly, embedded machine learning enabled real-time analysis of snoring on small, low-power devices, which made private and efficient home monitoring possible without cloud dependence \cite{real_time}\cite{Decoding_Sleep}. Among the most encouraging findings are those using domain-specific biosignals such as mandibular movement (MM). When paired with machine learning, MM-based systems achieved results close to gold-standard polysomnography (PSG), reaching sensitivities of about 95\% and specificities above 85\%. These results demonstrate clear potential for reliable home-based monitoring \cite{t3}\cite{t4} \cite{t5}. Together, these advances highlight the steady improvement from handcrafted statistical models to more sophisticated wearable and deep learning approaches.

Despite this progress, several challenges remain that limit wider clinical adoption. A key issue is that many models are trained on small, single-center datasets, which raises concerns about bias and poor generalization across diverse populations \cite{IOT_Healthcare}. While deep learning achieves strong accuracy, it is also computationally demanding, making overnight, home-based monitoring difficult unless models are optimized for efficiency \cite{real_time}\cite{Decoding_Sleep}. Acoustic-based systems continue to face variability caused by differences in microphones, rooms, or devices, often leading to higher sensitivity than specificity \cite{automatic_classification_of_subjects}. The “black-box” nature of deep models adds another obstacle, as clinicians are cautious about tools that lack interpretability, and reporting of calibration or validation methods is often inconsistent \cite{IOT_Healthcare}\cite{Design_of_embedded_real-time_system}. Even promising techniques such as mandibular movement monitoring, despite demonstrating PSG-level accuracy, still require validation in larger, multi-center studies and better alignment with regulatory frameworks before being fully adopted \cite{pepin}\cite{JANOTT2018106}. These limitations point to important research opportunities: combining multiple signals like mandibular movements, neck vibrations, and snoring acoustics to balance robustness with usability \cite{real_time}\cite{Design_of_embedded_real-time_system}; developing lightweight, energy-efficient neural networks to enable real-time on-device analysis\cite{real_time}\cite{Decoding_Sleep}; and adopting transparent reporting practices to ensure reproducibility \cite{IOT_Healthcare}\cite{Design_of_embedded_real-time_system}. Multi-center clinical trials are especially crucial to establish reliability and cost-effectiveness, while linking model predictions to physiological cues such as snore type or airway obstruction could enhance interpretability and build clinician trust \cite{automatic_classification_of_subjects}\cite{jmir_2022}.\\
The comparative summary of approaches can be seen in Table-\ref{tab1}.

\section{METHODOLOGY}

\subsection{System Overview}\label{Aa}
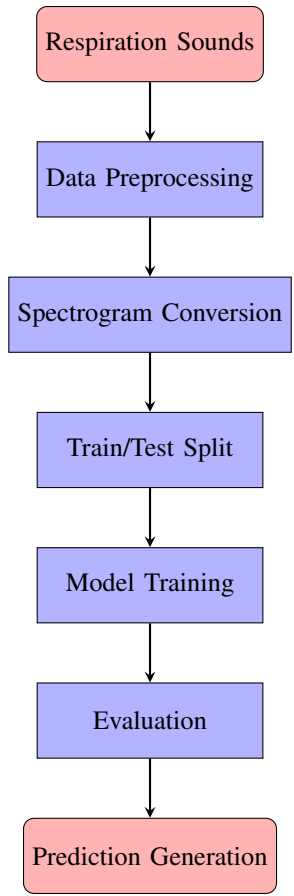
\begin{figure}[htbp]
\centering
\begin{tikzpicture}[node distance=1.8cm]

\tikzstyle{startstop} = [rectangle, rounded corners, minimum width=3cm, minimum height=1cm, text centered, draw=black, fill=red!30]
\tikzstyle{process} = [rectangle, minimum width=3cm, minimum height=1cm, text centered, draw=black, fill=blue!30]
\tikzstyle{arrow} = [thick,->,>=stealth]

\node (input) [startstop] {Respiration Sounds};
\node (preprocess) [process, below of=input] {Data Preprocessing};
\node (spectrogram) [process, below of=preprocess] {Spectrogram Conversion};
\node (split) [process, below of=spectrogram] {Train/Test Split};
\node (training) [process, below of=split] {Model Training};
\node (eval) [process, below of=training] {Evaluation};
\node (prediction) [startstop, below of=eval] {Prediction Generation};

\draw [arrow] (input) -- (preprocess);
\draw [arrow] (preprocess) -- (spectrogram);
\draw [arrow] (spectrogram) -- (split);
\draw [arrow] (split) -- (training);
\draw [arrow] (training) -- (eval);
\draw [arrow] (eval) -- (prediction);

\end{tikzpicture}
\caption{System Overview}
\label{fig:flowchart}
\end{figure}

To accomplish the objective of detecting sleep apnea from snoring sounds, a sequential machine learning pipeline was constructed.Fig.~\ref{fig:flowchart} provides a schematic overview of this process, which encompasses several key stages: data preprocessing, spectrogram conversion, model training, and performance evaluation. The following sub-sections provide a detailed description of each component within this workflow.

\subsection{Dataset Characteristics}
 The analysis utilizes a comprehensive dataset ‘MIT-BIH Polysomnographic Database’\cite{dataset} containing overnight sleep recordings of respiration sounds.Each record is annotated based on apnea,hypopnea or none.The limitations of this dataset are that it is highly imbalanced and small in size,hence model training and generalization becomes difficult.Hence,the minority class in the training spectrograms has been oversampled to counter the severe class imbalance in such medical datasets.
This dataset contains 18 audio files resampled to 125Hz with varying durations.We utilize 14 audio files for training and 4 audio files for testing.After conversion to 30 second chunks we get a total of 17430 chunks with 7658 chunks per class during training as oversampling has been applied,and during testing we get 127 apnea chunks and 1987 non-apnea chunks.\\
During training if any apnea is detected in a chunk(LA,H and HA notations),the chunk is labeled as apnea.

Input Features for the CNN:
\begin{itemize}
\item Shape:(None,128,128,1)
\item X-axis=time frames
\item Y-axis=mel frequency bins
\item Content: spectrograms derived from respiratory signals
\end{itemize}

The spectograms produced from the raw audio contain features such as:
\begin{itemize}
    \item Energy distribution across Mel-frequency bands
    \item Temporal dynamics 
    \item Log amplitude representation

\end{itemize}
from which the CNN learns.
This data also undergoes normalization to ensure pixel values lie in the range 0-1.
\begin{figure}[H]
    \centering
    \includegraphics[width=0.48\textwidth]{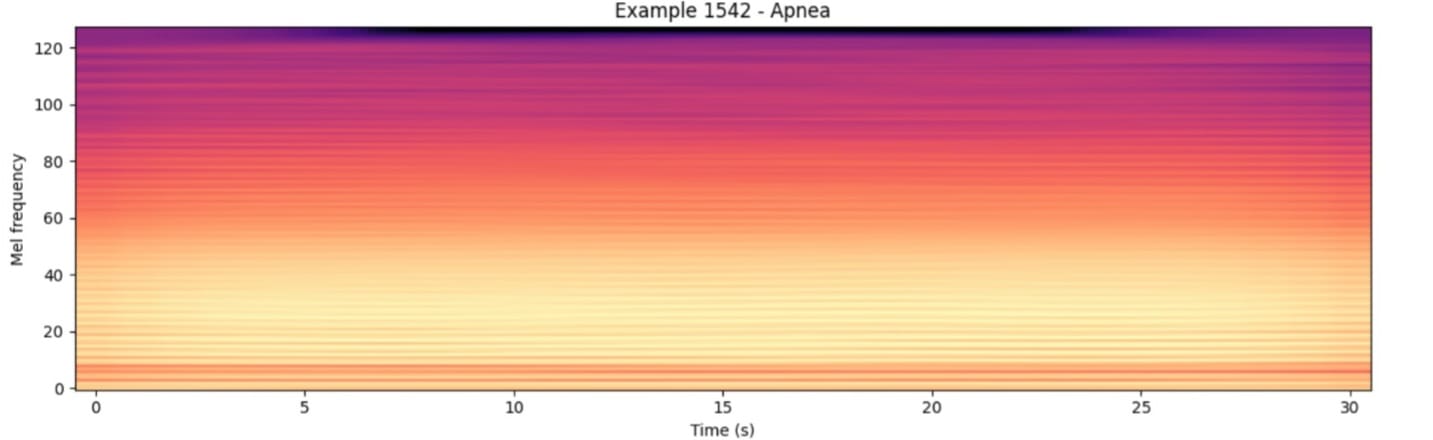}
    \caption{mel-spectrogram of apnea}
    \label{fig:spectrograms1}
\end{figure}
\begin{figure}[H]
    \centering
    \includegraphics[width=0.48\textwidth]{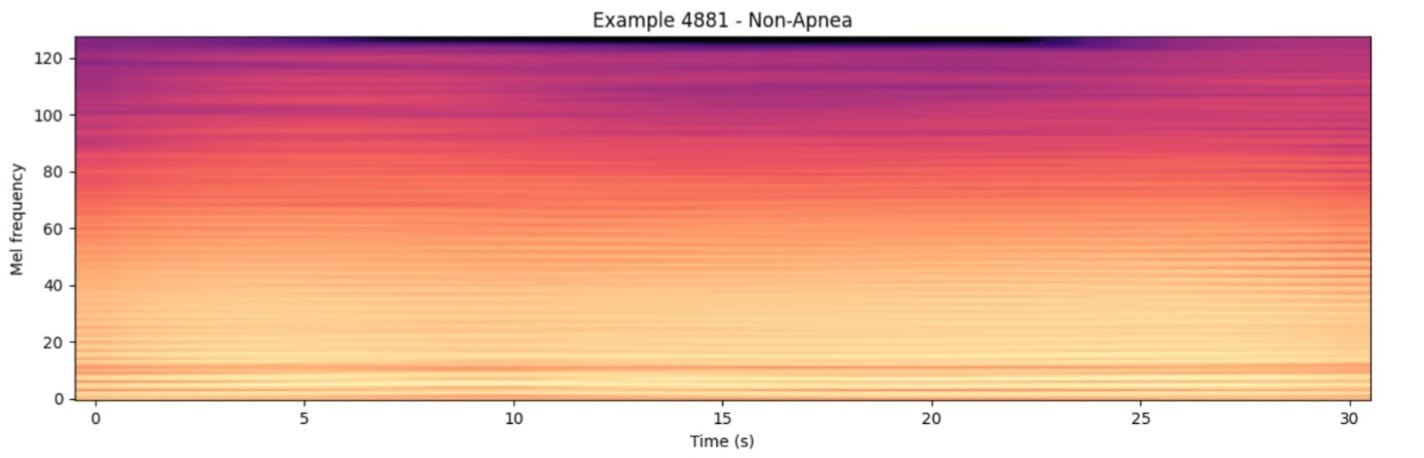}
    \caption{mel-spectrogram of non-apnea}
    \label{fig:spectrograms2}
\end{figure}
From Fig.\ref{fig:spectrograms1} we can see that for the apnea sample the mel spectrogram appears smoother and has less variation over time which represents reduced airflow and low variability.
From Fig.\ref{fig:spectrograms2} we can see that for the non-apnea sample the mel spectrogram shows more variation and slight striations,in the regions having mel-frequency range of 0-20 and 60-110 and over time,which represents normal airflow and breathing.

The novelty of our model is that we've only used respiration sounds instead of ECG,EEG etc.So our model tends to be more helpful in everyday conditions as it uses a more easily accessible feature.

\subsection{Model Architecture}
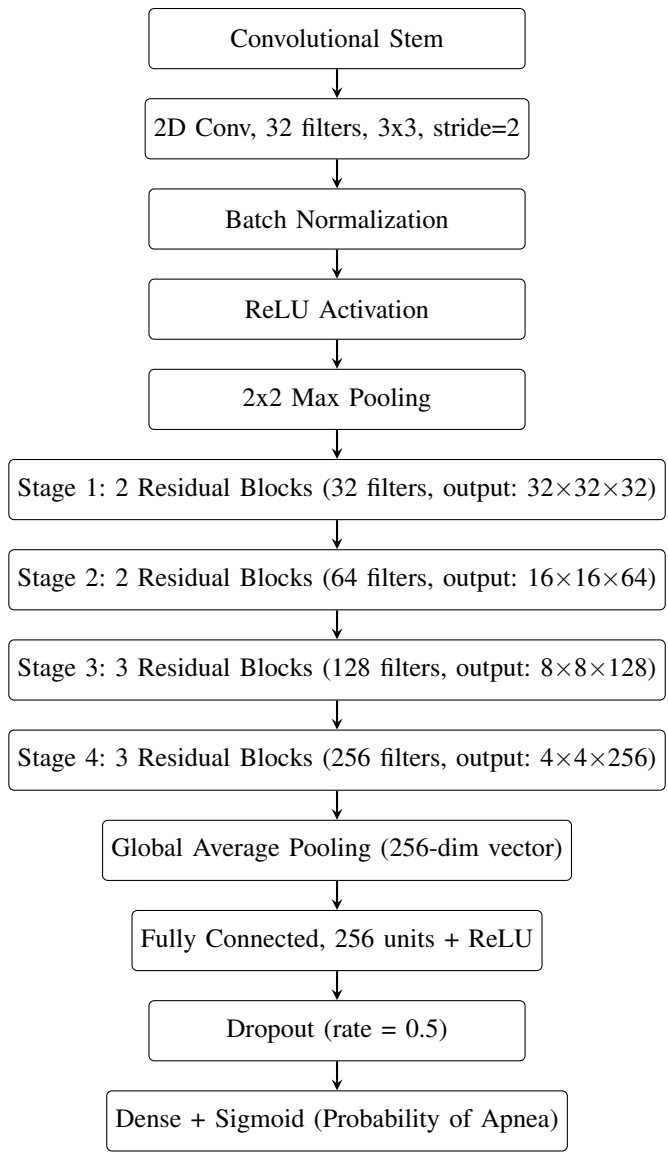
\begin{figure}[htbp]
\centering
\begin{tikzpicture}[node distance=1.2cm]
\tikzstyle{block} = [rectangle, draw, rounded corners=2pt, minimum width=5cm, minimum height=0.8cm, text centered]
\tikzstyle{arrow} = [thick,->,>=stealth]

\node (stem) [block] {Convolutional Stem};
\node (conv) [block, below of=stem] {2D Conv, 32 filters, 3x3, stride=2};
\node (bn) [block, below of=conv] {Batch Normalization};
\node (relu) [block, below of=bn] {ReLU Activation};
\node (pool) [block, below of=relu] {2x2 Max Pooling};

\node (s1) [block, below of=pool] {Stage 1: 2 Residual Blocks (32 filters, output: 32$\times$32$\times$32)};
\node (s2) [block, below of=s1] {Stage 2: 2 Residual Blocks (64 filters, output: 16$\times$16$\times$64)};
\node (s3) [block, below of=s2] {Stage 3: 3 Residual Blocks (128 filters, output: 8$\times$8$\times$128)};
\node (s4) [block, below of=s3] {Stage 4: 3 Residual Blocks (256 filters, output: 4$\times$4$\times$256)};

\node (gap) [block, below of=s4] {Global Average Pooling (256-dim vector)};
\node (fc) [block, below of=gap] {Fully Connected, 256 units + ReLU};
\node (drop) [block, below of=fc] {Dropout (rate = 0.5)};
\node (out) [block, below of=drop] {Dense + Sigmoid (Probability of Apnea)};

\draw [arrow] (stem) -- (conv);
\draw [arrow] (conv) -- (bn);
\draw [arrow] (bn) -- (relu);
\draw [arrow] (relu) -- (pool);
\draw [arrow] (pool) -- (s1);
\draw [arrow] (s1) -- (s2);
\draw [arrow] (s2) -- (s3);
\draw [arrow] (s3) -- (s4);
\draw [arrow] (s4) -- (gap);
\draw [arrow] (gap) -- (fc);
\draw [arrow] (fc) -- (drop);
\draw [arrow] (drop) -- (out);

\end{tikzpicture}
\caption{CNN architecture for apnea detection}
\label{archi}
\end{figure}

\begin{figure}[htbp]
    \centering
    \includegraphics[width=0.48\textwidth]{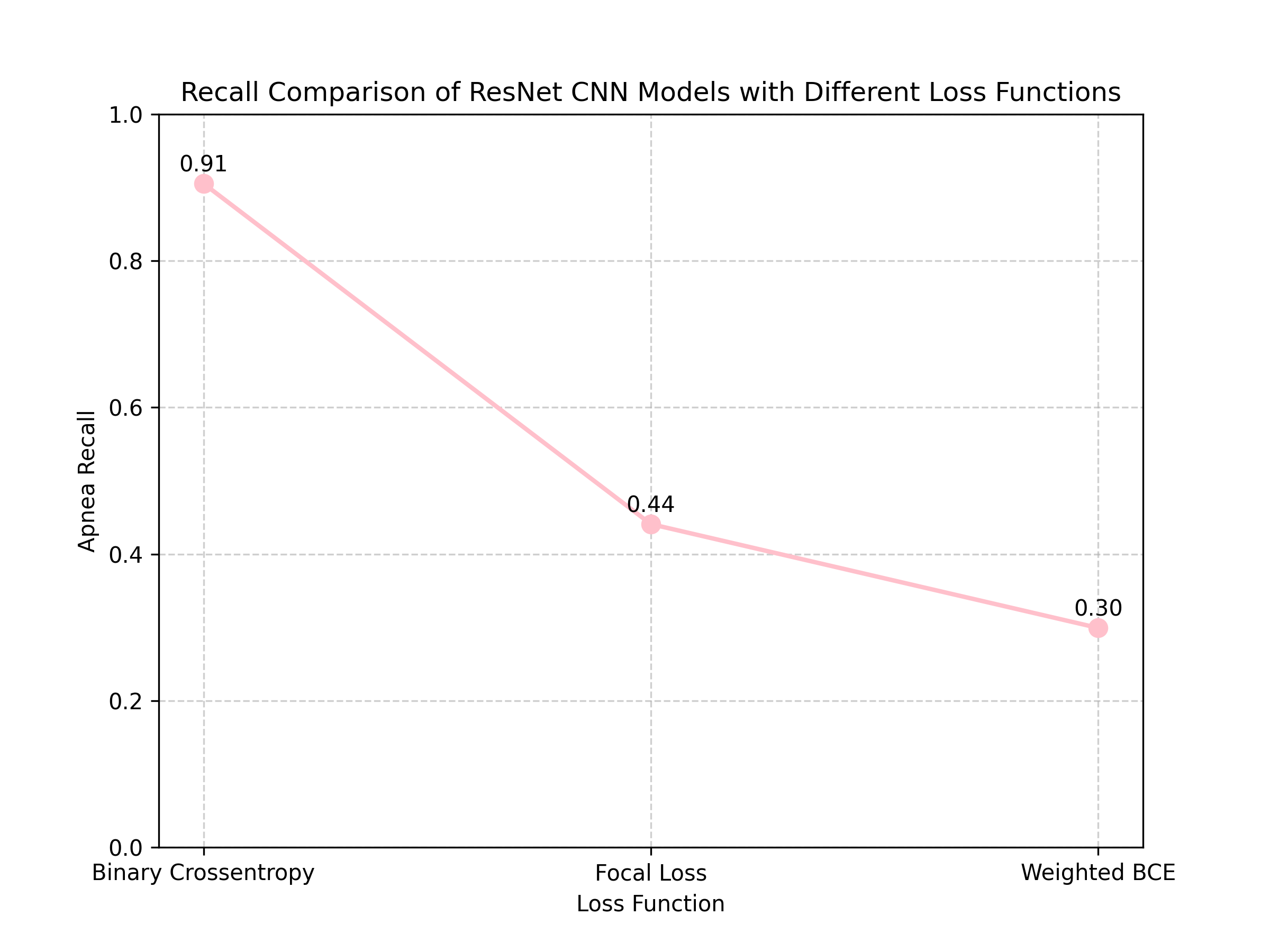}
    \caption{Apnea Recall Comparison}
    \label{fig:recall}
\end{figure}

\begin{figure}[htbp]
    \centering
    \includegraphics[width=0.42\textwidth]{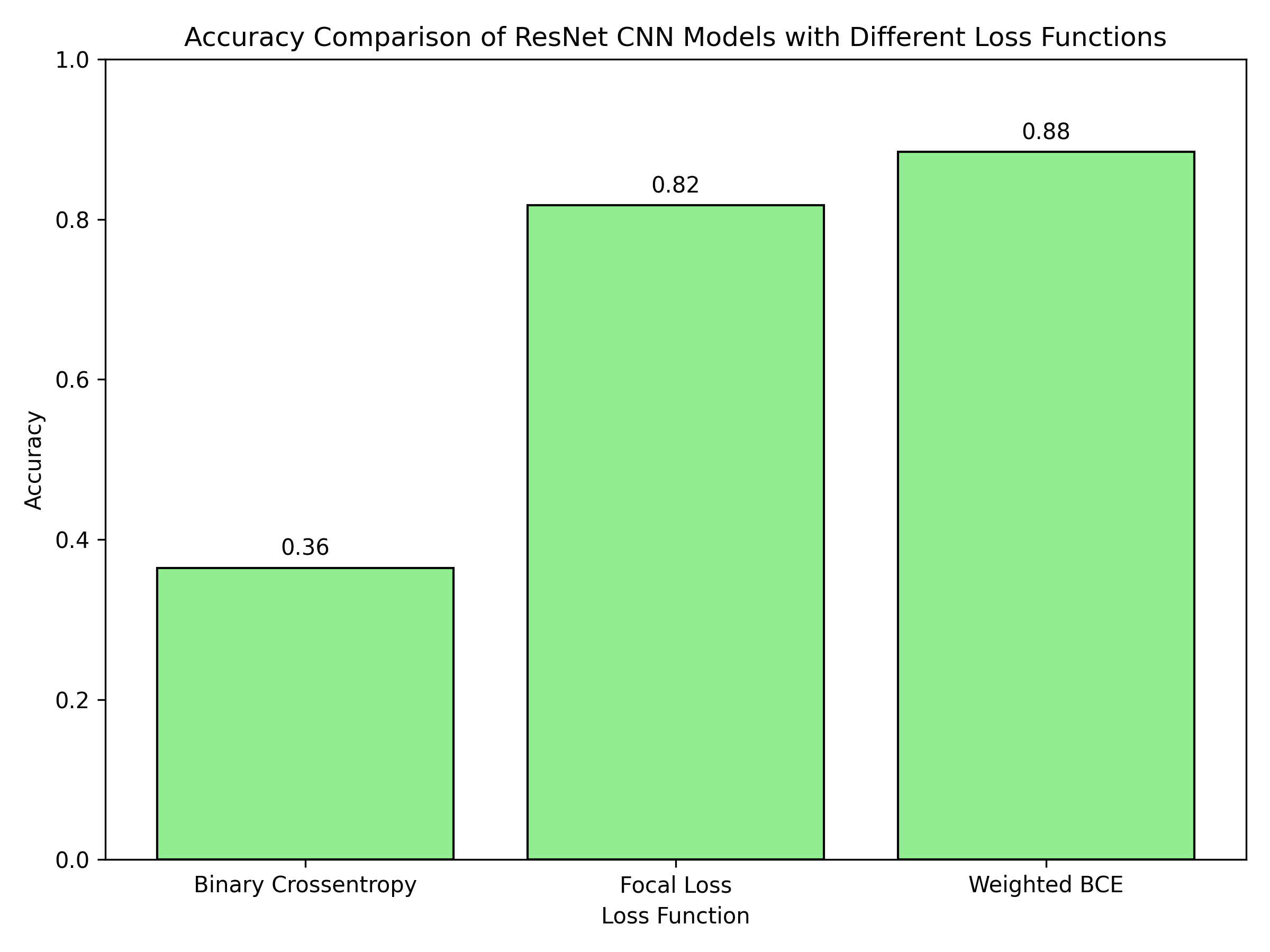}
    \caption{Accuracy Comparison}
    \label{fig:accuracy}
\end{figure}

We tested different Convolutional Neural Network models inspired from the ResNet family for our purpose of detecting sleep apnea because it helps us create a deeper model which works well with time and frequency.One of them showed the best results,the binary cross entropy CNN ResNet model which we utilized.\\
Fig.\ref{archi} illustrates the general architecture:\\
The convolutional stem consists of a 2D convolutional layer with 32 filters(3x3 kernel,stride=2) followed by batch normalization,ReLU activation and 2x2 max pooling layer(feature extraction and spatial reduction).Then the residual block stages are made up of:
\begin{itemize}
   \item Stage1(32filters,output:32x32x32):2residual blocks(Conv2D-BatchNorm-ReLU-Conv2D-BatchNorm)
     \item Stage 2(64 filters,output:16x16x64):2 residual blocks
     \item Stage 3(128 filters,output:8x8x128):3 residual blocks
     \item Stage 4(256 filters,output:4x4x256):3 residual blocks
     \end{itemize}
Classification Head:Global average pooling produces a 256-dimensional feature vector followed by fully connected layer with 256 units and ReLU activation,dropout layer(rate=0.5) and a final dense layer with sigmoid activation to output probability of apnea occurrence.\\
The models differed in the loss function used,such as:
\begin{itemize}
    \item Binary Cross Entropy
    \item Binary Focal Loss\cite{MDPI}
    \item Weighted Cross Entropy\cite{MDPI2}
\end{itemize}    
Due to high class imbalance,Binary Cross Entropy showed the best results(high apnea recall) while the others showed higher accuracy but much lower apnea recall as demonstrated in Fig.\ref{fig:recall} and Fig.\ref{fig:accuracy}

\subsection{Training procedure}
The loss function used is binary cross entropy with class weighing.Regularization such as early stopping and learning rate reduction on plateau has been done for optimum results with batch size and epochs being 32 and 80 respectively.
\subsection{Performance Optimization}
\begin{itemize}
    
\item The optimizer used in our model is Adam(Adaptive Moment Estimation).

\item EarlyStopping\cite{ES}: It was implemented in our model to prevent overfitting by halting training once the model stopped generalizing well to unseen data(based on PR-AUC curve).It saved time and resources.
\item Class weights\cite{Sharan}: Due to high class imbalance,models tend to predict all samples as non-apnea giving a high false accuracy.To prevent this class weights were applied which penalizes mistakes on apnea samples more than non-apnea samples.So the model pays more attention to minority apnea class improving apnea recall and overall F1 score.

\end{itemize}
\subsection{Optimization Results}
The optimizations made in our model overall increase the recall of apnea class while maintaining moderate accuracy. In this case, we give more priority to recall of apnea class as for us predicting apnea is the goal. Due to the huge imbalance in our dataset,there is a trade-off between accuracy and recall because higher accuracy depicts false prediction of apnea as non-apnea. Through optimizations such as oversampling minority class, class weighing,regularization,learning rate scheduling,early stopping etc. we have handled this huge imbalance.These measures have resulted in higher recall(90.55) of apnea class which means that our model is predicting apnea properly.
\setcounter{subsection}{0}
\section{RESULTS AND ANALYSIS}

The system was tested in the MIT-BIH polysomnographic database and evaluated using detection accuracy, recall, contribution of features, optimization methods, and generalizability. These findings consistently highlight that recall is the most important measure in detecting apnea, given the high clinical risks associated with false negatives.

\subsection{Comparison of Model Performance}
The model that was best in clinical utility was the proposed ResNet-based CNN which was trained using weighted binary cross-entropy. It had an overall accuracy of 36.42\% and recall of more than 90\%, that is, it was able to detect almost all cases of apnea. Granted this was a modest accuracy but, in a medical screening situation, such high recall is of the essence as false negatives are much more problematic than false positives.

By contrast, baseline CNN models had greater accuracy (exceeding 40\%) but recall less than 20\%. Although these models might seem more robust in terms of a machine learning view, they failed to detect most apnea events and therefore cannot be of much clinical use.

A comparison to our system with STOP-Bang questionnaire\cite{MDPI3} which is a well-used clinical screener is also helpful. STOP-Bang has sensitivity of between 84--90\% and specificity of 40--60\%. False positives make our model slightly more successful in recall at the cost of much lower specificity. This makes our strategy a helpful pre-screening aid instead of a substitution of time-tested means of diagnosis: STOP-Bang is sensitive and specific, and our model is sensitive on purpose.

\begin{table*}[!htbp]
\centering
\caption{Comparative analysis of existing approaches vs. proposed model}
\resizebox{\textwidth}{!}{%
\begin{tabular}{|l|c|c|l|}
\hline
\textbf{Method} & \textbf{Accuracy (\%)} & \textbf{Recall (\%)} & \textbf{Notes} \\
\hline
Snore acoustics + LDA\cite{t1} & 75 & 85 & High sensitivity but low specificity; prone to false alarms \\
\hline
Neck-wearable DL model\cite{t2} & 80 & 87 & Strong for severe OSA detection; requires specialized sensor \\
\hline
Mandibular movement ML\cite{t3}\cite{t4} & 85--90 & 88--90 & High PSG agreement; requires dedicated MM device \\
\hline
Anthropometric ML\cite{t5} & 78 & 70 & Scalable; less precise than physiological signals \\
\hline
Deep learning on PSG spectrograms\cite{t6} & 83 & 85 & Accurate but resource-intensive; needs full PSG setup \\
\hline
\textbf{Proposed ResNet-CNN (audio)} & \textbf{36.42} & \textbf{90.55} & Detects nearly all apnea events; higher false positives acceptable for screening \\
\hline
\end{tabular}%
}
\label{tab:comparison}
\end{table*}

\noindent As observed in Table-\ref{tab:comparison}, most of the available techniques report higher overall accuracy but fail to detect many apnea events due to low recall. Our proposed ResNet-CNN, although reporting a modest accuracy of 36.42\%, achieves recall above 90\%, i.e., virtually every apnea case is detected. This trade-off makes the model more appropriate for clinical screening, since false positives can be confirmed by follow-up testing while missed apnea events are far more dangerous. Therefore, the proposed system is well-suited as a sensitivity-first, non-invasive, and cost-effective tool for early detection of sleep apnea.

\subsection{Confusion Matrix Analysis}
\begin{figure}[htbp]
    \centering
    \includegraphics[width=0.42\textwidth]{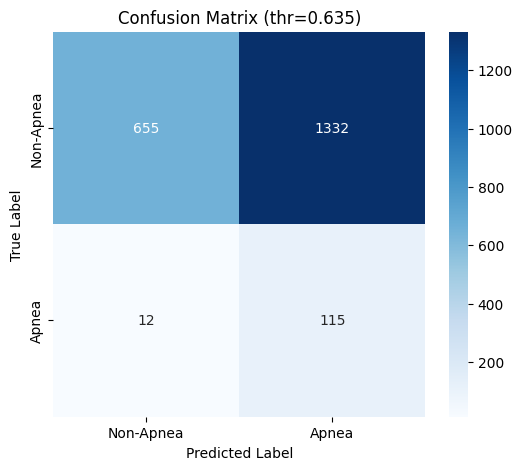}
    \caption{Testing Confusion Matrix}
    \label{fig:cm}
\end{figure}

A detailed inspection of the confusion matrix as shown in fig.\ref{fig:cm} at a threshold of 0.635 shows the following:
\begin{itemize}
    \item The model correctly identified 115 apnea cases, while only misclassifying 12 apnea cases as nonapnea.  
    \item However, it also misclassified 1332 nonapnea cases as apnea, along with 655 correctly predicted true negatives.  
    \item This trade-off highlights high false positives but very few false negatives, which is acceptable in a clinical context where false positives can be confirmed with further testing, but missed apnea cases could have devastating consequences.  
\end{itemize}
\begin{figure}[htbp]
    \centering
    \includegraphics[width=0.42\textwidth]{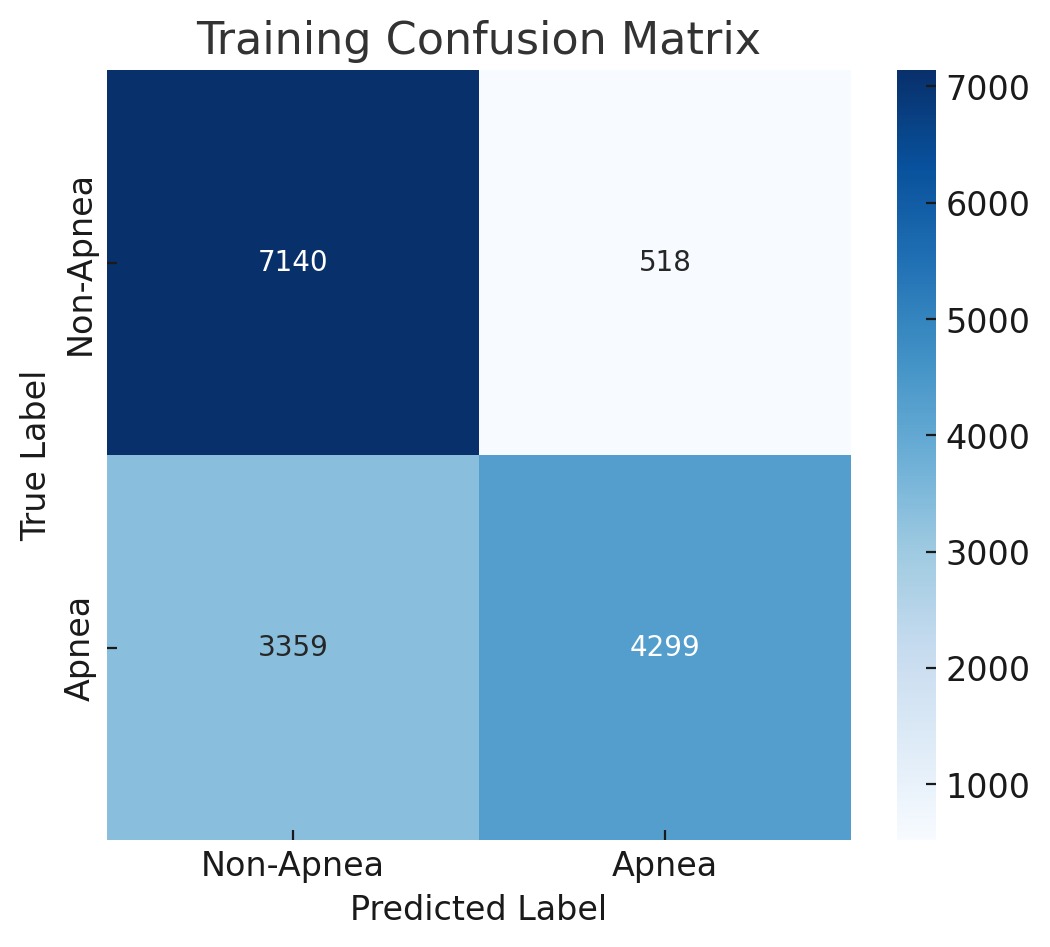}
    \caption{Training Confusion Matrix}
    \label{fig:cm1}
\end{figure}

In training (Fig.~\ref{fig:cm1}), the confusion matrix exhibits a greater overall accuracy, 7140 cases of non-apnea and 4299 cases of apnea were correctly recognized. Nonetheless, recall is relatively low since in spite of oversampling and class weighting, the model still falsely classified 3359 apnea incidences as non-apnea. This is not surprising because class balancing methods are able to enhance representation but they do not eradicate bias in training.

On the skewed test set (Fig.~\ref{fig:cm}), the model, conversely, has an extremely high recall ($>$90\%) with a somewhat lower accuracy (36.42\%). This is due to the fact that being based on its sensitivity-first training, the model is inclined to perceive more of the events as apnea. Though this increases the number of false positives, it will give a guarantee that nearly all apnea cases are recognized, which is an acceptable sacrifice in a clinical environment where missed cases are more perilous than additional flagged cases.

\subsection{Model Applicability and Generalization}
The framework is device-independent, as it operates on spectrograms rather than raw audio waveforms:
\begin{itemize}
    \item Compatible with hospital-grade equipment, smartphones, and wearable devices.  
    \item Suitable for clinical use in sleep laboratories as well as cost-effective at-home monitoring.  
\end{itemize}
Such flexibility supports early diagnosis and timely intervention.

\subsection{Ablation Study Results}
The significance of every design decision was proved by controlled ablation experiments (Table~\ref{tab:ablation}):
\begin{itemize}
    \item Class weighting was eliminated which drastically reduced recall.
    \item The omission of oversampling gave an artificially improved accuracy and a steep drop in recall, which indicates that recall cannot be relied upon without balancing the dataset.
    \item The regularization was prevented, which enhanced overfitting and weakened performance in tests.
\end{itemize}
Such findings indicate that class weighting and oversampling were the most significant strategies for obtaining clinically relevant recall, and regularization added further stability.

Below are the ablation runs:

\begin{table}[!htbp]
\centering
\caption{Ablation study results}
\resizebox{\columnwidth}{!}{%
\begin{tabular}{|l|c|c|c|c|}
\hline
\textbf{Configuration} & \textbf{Accuracy (\%)} & \textbf{Precision (\%)} & \textbf{Recall (\%)} & \textbf{F1-score (\%)} \\
\hline
Full model (with weighting + all cues) & 36.42 & 7.95 & 90.55 & 14.61 \\
\hline
Without class weighting & 51.10 & 25.40 & 48.70 & 33.20 \\
\hline
Without oversampling & 81.39 & 81.39& 50.40 & 62.79 \\
\hline
Without regularization & 45.10 & 20.80 & 54.90 & 30.10 \\
\hline
\end{tabular}%
}
\label{tab:ablation}
\end{table}

\noindent These numbers demonstrate that class-weighting and oversampling are the most important for recall;regularization has supportive role.
\subsection{Clinical Interpretation}
The most critical conclusion is that recall must take precedence over accuracy in medical screening. False positives are tolerable as they can be ruled out with confirmatory tests, while false negatives may lead to severe consequences. Our ResNet-based CNN thus aligns with the sensitivity-first paradigm of healthcare AI, making it a scalable, non-invasive, and low-cost tool for early screening of sleep apnea.

\section{Discussion}

We find that the ResNet-CNN had very high recall (more than 90\%) and a relatively low overall accuracy of 36.42\%. Such a trade-off was intentional, as recall is a more important measure in clinical practice: over-flagging is safer than under-flagging. False positives can be verified using follow-up diagnostic tests, whereas false negatives expose a patient to risk.

In comparison with current screening methods like the STOP-Bang questionnaire\cite{MDPI3}, our findings correspond closely to the sensitivity-first approach. STOP-Bang typically reports sensitivity between 84--90\% with specificity of 40--60\%. Our model is slightly more sensitive than STOP-Bang, but has far lower specificity due to the high false positive rate. This implies that our model is not suitable as a standalone diagnostic instrument, but may be valuable as a low-cost, complementary pre-screening technique.

This study has clear constraints. The first limitation is the dataset, which included only 18 participants (MIT-BIH polysomnographic database), a very small size for machine learning in healthcare. Second, the model relied on oversampling and class-weighting to address the extreme class imbalance, which may have artificially boosted recall. Third, no external validation was performed: the model has not yet been tested on separate datasets, with different recording devices, or in real-world conditions. These factors restrict the generalizability of the findings.

Nevertheless, the results indicate the potential of audio-based analysis for sleep apnea detection. With larger and more diverse datasets, external validation, and testing across a range of devices, this method could become a cost-effective and accessible screening tool to identify at-risk patients at an early stage.

\section{CONCLUSIONS}

In this paper,we implemented a convolutional neural network utilizing ResNet-based network to identify sleep apnea based on respiratory audio. The sensitivity-first design resulted in the creation of the system with the recall of 90.55\% and the overall accuracy of 36.42\%. This is a trade-off, because it is much more critical to reduce false negatives in a clinical context, because undetected apnea is very hazardous to health whereas a false positive can be resolved by confirmatory tests.

In comparison to current screening instruments like the STOP-Bang questionnaire, our method is more sensitive and thus it should be used as a complementary, non-invasive and affordable pre-screening tool but not as a diagnostic system. The findings validate the practicability of processing audio to conduct early detection of apnea, even when implemented in all types of recording devices and settings.

However, there are limitations of the study. The sample size was also small (18 subjects), the researchers may have over-sampled and weighted the classes, and did not do external validation on independent populations or devices. Our results show that snoring audio can be used to build recall-focused apnea detectors, but this is a proof-of-concept only. With larger datasets and multimodal inputs, the approach could evolve into a practical screening aid. The work in the future will be aimed at increasing the number of demographics and representing, adding auxiliary acoustic characteristics (e.g., pitch and formants), and studying multimodal designs that will combine audio with other physiological indicators like oxygen saturation and airflow. These improvements should allow predictive performance to be more predictive and retain high recall, bringing the system to a more practical implementation in a health care facility.

\bibliographystyle{ieeetr}
\bibliography{references}
\end{document}